\documentclass[
    ,final            
  ]
  {aipproc}

\layoutstyle{8x11double}

\begin{document}

\title{A cosmological AMR MHD module for Enzo}


\author{Hao Xu}{
  address={Department of Physics and CASS, University of California, San Diego, San Diego, CA}
  ,altaddress={Theoretical Astrophysics, Los Alamos National Laboratory, Los Alamos, NM} 
}

\author{David C. Collins}{
  address={Department of Physics and CASS, University of California, San Diego, San Diego, CA}
}

\author{Michael L. Norman}{
  address={Department of Physics and CASS, University of California, San Diego, San Diego, CA}
}

\author{Shengtai Li}{
  address={Mathematical Modeling and Analysis, Los Alamos National Laboratory, Los Alamos, NM}
}

\author{Hui Li}{
  address={Theoretical Astrophysics, Los Alamos National Laboratory, Los Alamos, NM}
}

\begin{abstract}
 Magnetic fields play an important role in almost all astrophysical
phenomena including
star formation. But due to the difficulty in analytic modeling and
observation,
magnetic fields are still poorly studied and numerical simulation has
become a major tool. We have implemented a cosmological magnetohydrodynamics
package for Enzo which is an AMR hydrodynamics code designed to simulate
structure formation. We use the TVD solver developed by
S. Li as the base solver. In addition, we employ the constrained transport (CT)
algorithm as described by D. Balsara. For interpolation magnetic fields to fine grids we
used a divergence free quadratic
reconstruction, also described by Balsara. We present results from several test
problems including MHD caustics,
MHD pancake and galaxy cluster formation with magnetic fields. We also discuss possible
applications of our AMR MHD code to first star research.
\end{abstract}

\maketitle


\section{Introduction}

  Adaptive mesh refinement(AMR) cosmological hydrodynamics simulations play an important role in the study of structure formation of different scales from galaxy clusters to first stars in the past ten years. Its ability to achieve very high resolution in large scale simulations with relatively small computer resources has helped us to understand the first stars formed in our Universe. The possible effects of magnetic fields have been largely ignored. It is well established that magnetic fields are present on different scales, from intracluster medium to interstellar medium. The origin and evolution of these magnetic fields and their role on the structure formation are still unclear. So, it is imperative to include magnetic fields into current hydrodynamics AMR cosmology code. In this paper, we present the newly developed MHD version of the Enzo, which is wildly used in the study of first stars \citep{Abel02, O'Shea05, O'Shea07}.              

\section{MHD with ENZO}

The MHD equations in the comoving coordinates are:

\begin{eqnarray}{}
\frac{\partial \rho}{\partial t} + \frac{1}{a} \nabla \cdot ( \rho \bf{v}) & = & 0 \\
\frac{\partial \rho \bf{v}}{\partial t} + \frac{1}{a} \nabla \cdot (\rho \bf{v}\bf{v} + \bar{p} - \bf{B}\bf{B}) & = & - \frac{\dot{a}}{a} \rho \bf{v}- \frac{1}{a}\rho \nabla \phi  \\
\frac{\partial E}{\partial t} + \frac{1}{a}\nabla \cdot [\bf{v}(\bar{p}+E)-\bf{B}(\bf{B}\cdot \bf{v})] & = & - \frac{\dot{a}}{a}(\rho v^{2}+3p+\frac{B^{2}}{2}) \nonumber \\  
& & - \frac{\rho}{a} \bf{v}\cdot \nabla \phi \\
\frac{\partial \bf{B}}{\partial t} - \frac{1}{a} \nabla \times (\bf{v} \times \bf{B}) & = &  -\frac{\dot{a}}{2a}\bf{B}
\end{eqnarray}
with
\begin{eqnarray}
E & = & \frac{1}{2} \rho v^2 + \frac{p}{\gamma-1} + \frac{1}{2}B^2 \\
\bar{p} & = & p + \frac{1}{2} B^2
\end{eqnarray}
where all variables have their usual meaning, a is the expansion parameter. To track the pressure more accurately in the supersonic region, we have also implemented the modified entropy equation given in \citet{Ryu93} and the internal energy equation given in \citet{Bryan95} in our code.

The MHD solver used for all the test problems here is a high-order Godunov-type finite-volume numerical solver developed by S.T. Li \citep{Li03, Li05}. This solver was recently successfully used to study magnetic jet problems \citep{Li06,Nakamura06,Nakamura07}.

We used a constrained transport(CT) scheme flux CT \citep{Balsara99} to maintain divergence-free magnetic fields. For the AMR hierarchy in Enzo, we used a modified divergence-free reconstruction scheme original proposed by \citet{Balsara01} including second order accurate divergence-free restriction and prolongation for magnetic fields. Details of the CT and AMR of magnetic fields in the MHD Enzo can be found in \citet{Collins07}. 

The MHD module has been tested and shown to be compatible with other physics packages installed in Enzo, such as radiative cooling, star formation and feedback. 

\section{Tests}

\subsection{MHD Caustic and Pancake}

\begin{figure}[h]
\includegraphics[width=.23\textwidth]{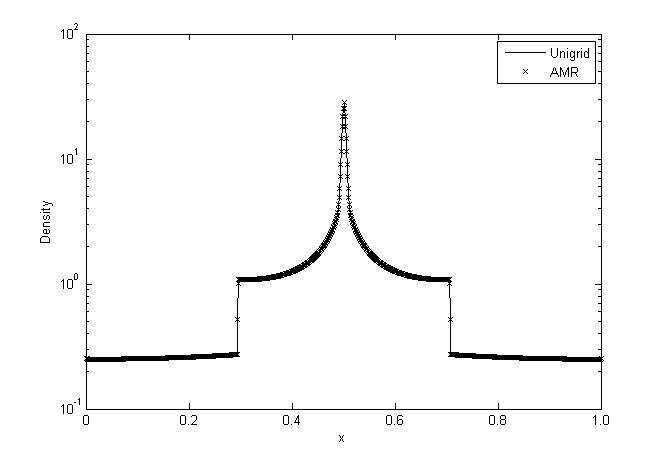}
\includegraphics[width=.23\textwidth]{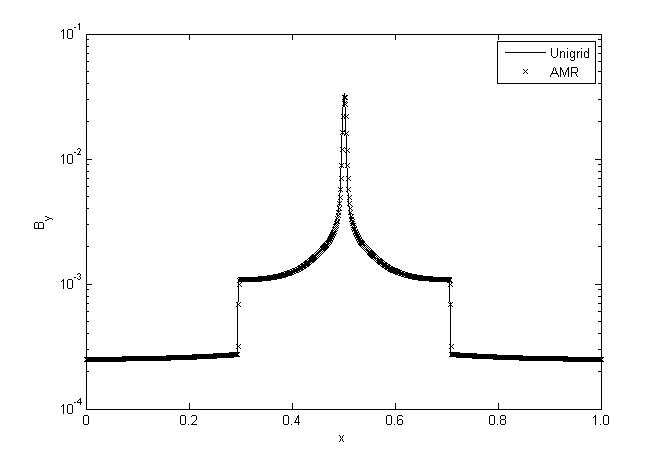}
\caption{Plots of density and y component of magnetic fields of MHD caustics at t=3. The initial magnetic field is $10^{-3}$.} 
\label{fig:caustic}
\end{figure}

The MHD Caustic test is taken from \citet{Li07} which generalizes the HD test of \citep{Ryu93}. The initial sinusoidal velocity field in the x-direction has the peak value $1/2\pi$, the initial density and pressure has been set to be uniform with $\rho=1$ and $p=10^{-10}$. Then caustics will be formed because of the compression by the velocity field. An initial uniform magnetic field of $10^{-3}$ in code units in the y direction was added to the simulation. Figure \ref{fig:caustic} compares the density and $B_y$ at $t=3$ of unigrid and AMR runs. AMR run had 256 cells in the root grid and 2 level refinements by 2. The AMR solution is indistinguishable from a uniform grid solution with 1024 cells.    

\begin{figure}[h]
\includegraphics[width=.23\textwidth]{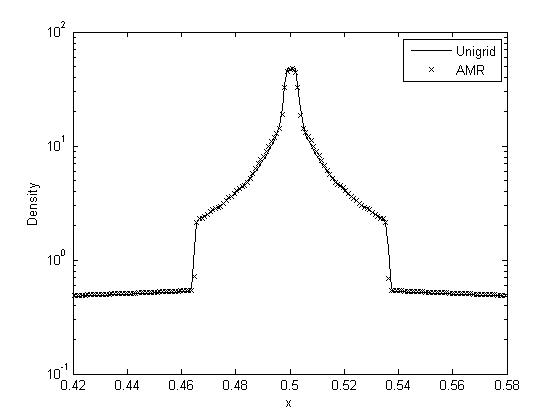}
\includegraphics[width=.23\textwidth]{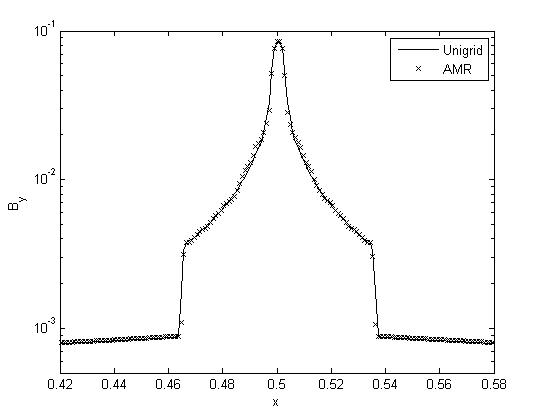}
\caption{Plots of density and y component of magnetic fields of MHD pancake.The initial magnetic field is $2 \times 10^{-5}G.$} 
\label{fig:pancake}
\end{figure}

Pancake is another standard test problem of cosmological hydrodynamics simulation \citep{Ryu93}. We have run the collapse of a one-dimension pancake in a purely baryonic universe with $\Omega=1$ and $h=\frac{1}{2}$. Initially, at $a_{i}=1$, which corresponds to $z_{i}=20$ in this test, a sinusoidal velocity field with the peak value $0.65/(1+z_{i})$ in the normalized unit has been imposed in a box with the comoving size $64h^{-1}Mpc$, so that the shock forms at $z=1$. The initial baryonic density and pressure have been set to be uniform with $\rho=1$ and $p=6.2\times 10^{-8}$ in the normalized code units. We applied initial uniform magnetic fields $B_y = 2.0 \times 10^{-5}G,~B_x=B_z=0$ in the simulation. We did the calculations with unigrid with 1024 cells and AMR run with 256 cells of root grid and 2 level refinements. Figure \ref{fig:pancake} shows the density and $B_y$ at $z=0$.  

\subsection{Cluster Formation with Magnetic Fields}

Cluster formation is one of the problems most widely studied by Enzo \citep{Norman05}. We have done this problem and compared with results from Enzo-ppm to test  our new code in large scale structure formation. The simulation is a $\Lambda$CDM
model with parameters $h=0.7$, $\Omega_{m}=0.3$, $\Omega_{b}=0.026$,
$\Omega_{\Lambda}=0.7$, $\sigma_{8}=0.928$. The survey volume is 256 $h^{-1}$ Mpc on
a side. The simulations were computed from a $128^3$ root grid and 2 level nested static grids in the Lagrangian region where the cluster forms which gives an effective root grid resolution of $512^3$ cells (0.5 $h^{-1}$ Mpc) and dark matter particles of mass $1.49 \times 10^{10} M_\odot$. AMR is allowed only in the region where the galaxy cluster forms, with a total of 8 levels of refinement beyond the root grid, for a maximum spatial resolution of 7.8125 $h^{-1}$kpc.  

\begin{figure}[h]
\includegraphics[width=.45\textwidth]{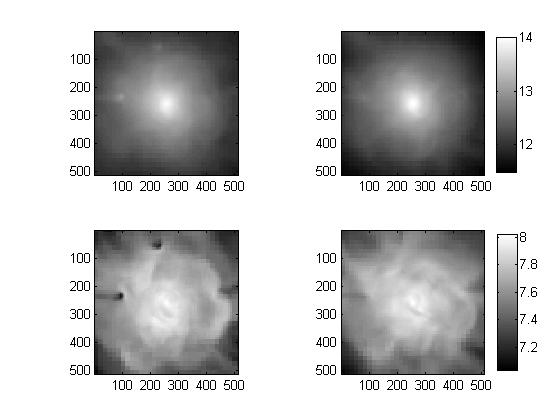} 
\caption{Logarithmic projected gas density(top) and logarithmic projected X-ray weighted temperature(down) at $z=0$ of adiabatic simulations. The images cover the inner 4 $h^{-1}$Mpc of cluster centers. The left panels show results from the PPM solver and the right panels show results from the MHD solver. The units of density and temperature are $M_\odot /Mpc^3$ and Kelvin respectively.} 
\label{fig:cluster2}
\end{figure}

We first present the results of no initial magnetic fields and compare them with the results from Enzo-ppm. The cluster parameters from the MHD code is almost identical to those from Enzo-ppm. The virial radii are 2.229Mpc from hydro and 2.226Mpc from MHD while the virial masses are $1.265 \times 10^{15} M_\odot$ for hydro and $1.260 \times 10^{15} M_\odot$ for MHD. Figure \ref{fig:cluster2} compare the projections of the baryon density and temperature.  

\begin{figure}[h]
\includegraphics[width=.45\textwidth]{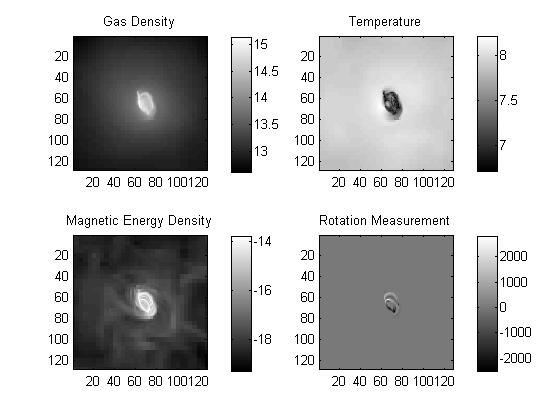}
\caption{Images of the baryon density (logarithmic, $M_\odot /Mpc^3$), temperature (logarithmic, Kelvin), magnetic energy density (logarithmic, $erg/cm^3$) and rotation measurement ($rad/m^2$) of the inner 1$h^{-1}$Mpc of cluster center. The initial magnetic field is $B_y = 1.0^{-9}G$ at $z=0$.} 
\label{fig:cluster4}
\end{figure}

We have performed simulation with initial magnetic fields, $B_x=B_z=0$, $B_y = 1.0^{-9}G$ with radiative cooling, star formation and star formation feedback. Figure \ref{fig:cluster4} shows the images of baryon density, temperature, magnetic energy density and Faraday rotation measurement of the cluster center. The rotation measurement is integral along the projection direction. 

\begin{figure}[h]
\includegraphics[width=.45\textwidth]{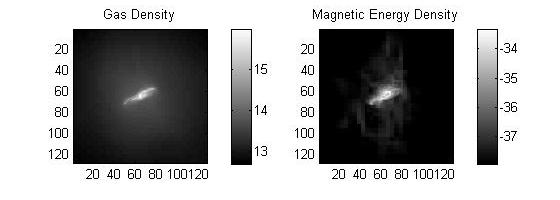}
\caption{Images of the baryon density (logarithmic, $M_\odot/Mpc^3$) and magnetic energy density (logarithmic, $erg/cm^3$) of the simulation with Biermann battery effect of the inner 1$h^{-1}$Mpc of cluster center. } 
\label{fig:cluster5}
\end{figure}

Another simulation we performed is without initial magnetic field but with the Biermann battery effect. The induction equation is modified by adding an additional battery term\citep{Kulsrud97}:

\begin{eqnarray}
\frac{\partial \bf{B}}{\partial t} & = & \nabla \times (\bf{v} \times \bf{B}) + \nabla \times (\frac{c \nabla p_e}{n_e e}) \\
                                   & = & \nabla \times (\bf{v} \times \bf{B}) + \frac{c m_H}{e} \frac{1}{1+\chi} \nabla \times (\frac{p}{\rho})
\end{eqnarray}
where c is speed of light, $p_e$ is the pressure of electron, $n_e$ is the electron number density, e is the electron charge, $m_H$ is the hydrogen mass and $\chi$ is the ionization fraction. We took $\chi = 1$ constant in space in our simulation. We performed this computation with radiative cooling. Figure \ref{fig:cluster5} shows the projection of logarithmic baryon density and the magnetic energy density of the cluster center.  

\section{Discussion}

We have introduced our MHD module for Enzo and presented some test results. With this new module, we have the ability to perform AMR MHD cosmology simulations. This module uses high accuracy TVD MHD solver and employs CT and AMR divergence-free magnetic fields reconstruction scheme to guarantee divergence-free of magnetic fields. MHD simulations using this MHD module can use all the exist physics packages in Enzo.    
 
It is widely believed that the magnetic fields should have little effects in the formation of the first stars, since maybe there were no magnetic fields at all. But even there are no magnetic fields from the early universe, the Biermann battery effect should have generated some very small fields during the collapse to form those stars. If the first generation stars spin very fast, these small seed fields could be maintained and amplified by dynamo effect. Then the magnetic fields in the first stars may pollute the environment by their explosion and play a part in the formation of next generation stars. This would be investigated in the near future.           


\begin{theacknowledgments}
This research was supported by IGPP at Los Alamos National Laboratory.
\end{theacknowledgments}




\begin{thebibliography}{9}

\bibitem[Abel et al.(2002)]{Abel02} Abel, T., Bryan, G. \& Norman, M.L.
2002, Science, 295, 93

\bibitem[Balsara(2001)]{Balsara01} Balsara, D. 2001, Journal of
Computational Physics, 174, 614

\bibitem[Balsara \& Spicer(1999)]{Balsara99} Balsara, D. \& Spicer, D., 1999,
 J. Comput. Phys., 149, 270

\bibitem[Bryan et al.(1995)]{Bryan95} Bryan, G. et al., 1995, Comp. Phys.,
89,  149

\bibitem[Collins et al.(2007)]{Collins07} Collins, D. et al. 2007, in preparation

\bibitem[Kulsrud et al.(1997)]{Kulsrud97} Kulsrud, R. M. et al., 1997, ApJ, 480, 481

\bibitem[Li \& Li(2003)]{Li03} Li, S \& Li, H. 2003, Technical Report, Los Alamos National Laboratory

\bibitem[Li(2005)]{Li05} Li, S., 2005 J. Comput. Phys., 203,344

\bibitem[Li et al.(2006)]{Li06} Li, H., Lapenta, G., Finn J.M., Li, S. \& Colgate, S. A., 2006, ApJ, 643, 92

\bibitem[Li et al.(2007)]{Li07} Li, S et al.,2007, ApJS, Accepted

\bibitem[Norman (2005)]{Norman05} Norman, M.L., 2005, Proc. Int. Sch. Phys. IOS, 1

\bibitem[Nakamura et al.(2006)]{Nakamura06} Nakamura, M., Li, H. \& Li, S., 2006, ApJ, 652, 1059

\bibitem[Nakamura et al.(2007)]{Nakamura07} Nakamura, M., Li, H. \& Li, S., 2007, ApJ, 656, 721

\bibitem[O'Shea et .al(2005)]{O'Shea05} O'Shea, B. et al., ApJ., 2005, 628, L5

\bibitem[O'Shea \& Norman(2007)]{O'Shea07} O'Shea, B. \& Norman, M.L., 2007, ApJ, 654,66 

\bibitem[Ryu et al.(1993)]{Ryu93} Ryu, D. et al.,1993, ApJ, 414, 1

\bibitem[Xu et al.(2007)]{Xu07} Xu, H. et al. in preparation

\end{thebibliography}


\end{document}